\documentclass[aps,prl,superscriptaddress]{revtex4-2}
\usepackage{lineno}
\usepackage{amsmath}
\usepackage{mathrsfs}
\usepackage{upgreek}
\usepackage{siunitx} 
\usepackage{graphicx}
\usepackage{multirow}
\usepackage{nameref}
\usepackage{amssymb}
\usepackage{color}
\usepackage{booktabs}
\usepackage{float}

\newcommand*{\dif}{\mathop{}\!\mathrm{d}}

\newcommand{\ve}[1]{\ensuremath{\mbox{\boldmath$#1$}}}
\newcommand{\mat}[1]{\ensuremath{\mathbb{#1}}}

\newcommand{\Rep}{\mathrm{Re_p}}
\newcommand{\phiswim}{\Phi_{\rm swim}}
\newcommand{\phisettle}{\Phi_{\rm settle}}

\newcommand{\massp}{m_{\mathrm{p}}}
\newcommand{\vp}{\ve v_{\mathrm{p}}}

\let\oldequation\equation
\let\oldendequation\endequation

\renewenvironment{equation}{\linenomathNonumbers\oldequation}{\oldendequation\endlinenomath}

\begin{document}
\title {Fluid inertial torque is an effective gyrotactic mechanism 
	for settling elongated micro-swimmers}
\author{Jingran Qiu}
\affiliation{AML, Department of Engineering Mechanics, Tsinghua University, 100084 Beijing, China}
\author{Zhiwen Cui} 
\affiliation{AML, Department of Engineering Mechanics, Tsinghua University, 100084 Beijing, China}
\author{Eric Climent}
\affiliation{Institut de M{\'e}canique des Fluides de Toulouse (IMFT), UMR5502 Universit{\'e} de Toulouse, CNRS. All{\'e}e du Prof. Camille Soula – 31400 Toulouse, France}
\author{Lihao Zhao}
\email{zhaolihao@mail.tsinghua.edu.cn}
\affiliation{AML, Department of Engineering Mechanics, Tsinghua University, 100084 Beijing, China}

\date{\today}

\begin{abstract}
Marine plankton are usually modeled as settling elongated micro-swimmers. For the first time, we consider the torque induced by fluid inertia on such swimmers, and we discover that they spontaneously swim in the direction opposite to gravity. 
We analyze the equilibrium orientation of swimmers in quiescent fluid and the mean orientation in turbulent flows using direct numerical simulations. 
Similar to well-known gyrotaxis mechanisms, the effect of fluid inertial torque can be quantified by an effective reorientation time scale. We show that the orientation of swimmers strongly depends on the reorientation time scale, and swimmers exhibit strong preferential alignment in upward direction when the time scale is of the same order of Kolmogorov time scale. Our findings suggest that the fluid inertial torque is a new mechanism of gyrotaxis that stabilizes the upward orientation of micro-swimmers such as plankton.

	\begin{description}
		\item[Keywords] swimmer, turbulence, gravitaxis, fluid inertia
	\end{description}

\end{abstract}

\maketitle

\section*{Introduction}
Plankton play an important role in marine ecosystem. For instance, plankton produce oxygen by photosynthesis, and transfer energy to zooplankton and other marine predators in the food web. Many motile plankton migrate vertically to pursue light or nutrients or to avoid predation \cite{smayda1997,hays2003review,bollens1989predator}. The vertical  migration is driven by gravity and other physical and chemical stimuli. The responses to these stimuli are known as gyrotaxis \cite{Pundyak2017}, phototaxis \cite{eggersdorfer1991phototaxis}, and chemotaxis \cite{stocker2008rapid}, etc.
Gyrotaxis is one of the important factors that influence the direction and efficiency of vertical migration~\cite{Pundyak2017}. 
Bottom-heaviness \cite{Kessler1986} and fore-aft asymmetry \cite{Roberts1970} are two well-known mechanisms of gyrotaxis. Many plankton are denser or wider at their rear parts than the front parts, and they are subjected to stabilizing torques due to gravity that reorients them in the upward direction.
Based on these two mechanisms, gyrotactic swimmers are widely studied by modeling them as point-wise motile particles that swim relative to the fluid under a gravitational torque~\cite{Lovecchio2019,Durham2009,Sengupta2017,Durham2013,DeLillo2014,Zhan2014,Gustavsson2016,Borgnino2018}.

Gyrotaxis causes swimmers to preferentially swim in vertical direction~\cite{Kessler1986,Zhan2014,fouxon2015phytoplankton,lewis2003orientation}, and to form spatial clustering \cite{Durham2013,Gustavsson2016,Borgnino2018,Durham2009}. The magnitude of gyrotactic torque is crucial to these phenomena, influencing not only the orientation but also the intensity and location of patchiness. To quantify gyrotaxis, it is important to identify possible mechanisms of gyrotaxis and quantify their contributions. However, question remains whether there exist other mechanisms for gyrotaxis. In particular, the widely used point-particle model neglects the effect of fluid inertial torque.

Recent studies indicated that the orientation of a spheroidal particle is affected by a fluid inertial torque \cite{Sheikh2020,Gustavsson2019}. This torque is a result of convective fluid inertial effect when a particle moves relative to the local fluid \cite{Dabade2015}.
Motile plankton are usually modeled as swimmers which move relative to the local fluid. The relative motion is due to their motility and the effect of gravitational settling, which results in a non-zero fluid inertial torque on settling micro-swimmers such as plankton.
Interestingly, we find that, elongated settling micro-swimmers reorient themselves in upward direction under the influence of fluid inertial torque. 
Therefore, we suggest that fluid inertial torque is a new mechanism of gyrotaxis, which is different from the two well-known mechanisms of bottom-heaviness and fore-aft shape asymmetry. 

In this paper, we introduce the model of settling swimmer, and analyze their orientation in both quiescent and turbulent flows. We show that the magnitude of fluid inertial gyrotaxis depends on the shape, the swimming and settling speeds of swimmers, and can be quantified by a dimensionless parameter that measures the time scale of reorientation under fluid inertial torque.

\section*{Fluid inertial torque on micro-swimmers}
\label{sec5}
\label{seciia}
Fluid inertial force and torque on a spheroid originate from the leading-order effects of fluid inertia when the spheroid moves relative to the fluid \cite{brenner1961oseen,Dabade2015}. The Reynolds number of a planktonic swimmer $\Rep=\left|\ve u - \ve v_{\rm p}\right| L/\gamma$ is usually much smaller than unity, where $\gamma$ is the kinematic viscosity of fluid. The small $\Rep$ is due to small size $L$ and weak motility of the swimmer that yields small velocity difference between the fluid $\ve{u}$ and the swimmer $\ve{v}_{\rm p}$. In the regime of $\Rep \ll1$, the fluid inertial correction for force is negligible because its magnitude is of the order of $\Rep$ and the swimmer experiences Stokes drag \cite{Sheikh2020,brenner1961oseen}.
However, the fluid inertial torque can be significant compared to the Jeffery torque \cite{Jeffery1922} that represents the effect of fluid velocity gradients on the rotation of swimmers. Ref.~\cite{Sheikh2020} showed that the ratio between the magnitudes of inertial torque and Jeffery torque in turbulence is proportional to $\left|\ve u-\ve v_{\rm p}\right|^2/u_{\eta}^2$ \cite{Sheikh2020}, where $u_{\eta}$ is the Kolmogorov velocity scale. Hence, the fluid inertial torque is not necessarily negligible when $\Rep \ll 1$, especially when a swimmer moves relative to the fluid at a significant speed. A detailed dimensional analysis is provided in Appendix A.

Plankton usually satisfy the overdamped limit, which means that the response time of their motion is much shorter than the characteristic timescale of fluid motion \cite{Gustavsson2019}.
In this case, plankton are usually modeled as point-wise spheroidal swimmers~\cite{Durham2009,Durham2013,Gustavsson2016,Lovecchio2019,Zhan2014}. The inertia of a swimmer is neglected, so its translational and rotational motion is governed by kinematic equations.
Following a similar approach in Ref.~\cite{Gustavsson2019}, we obtain the model of a settling micro-swimmer (Figure~\ref{figsketch}) with the influence of fluid inertial torque (see Appendix A for details).
The motion of a swimmer is governed by the following equations:
\begin{equation}
	\label{eqRotp}
	\dot{\ve n} = \ve {\omega}_{\rm p} \wedge \ve{n},
\end{equation}
\begin{equation}
	\label{eqVelp}
	\ve{v}_{\rm p} = \ve{u} + v_{\rm swim} \ve{n} + \ve{v}_{\rm settle}.
\end{equation}
Here,  $\ve{\omega}_{\rm p}$ is the angular velocity of the swimmer, and $\ve{n}$ is the unit vector along its symmetry axis. The swimmer is assumed to swim at a constant speed in the direction of $\ve{n}$, and it is advected by the local fluid with velocity $\ve{u}$. Settling due to gravity is taken into account by adding a settling speed $\ve{v}_{\rm settle}$. In the overdamped limit, a swimmer settling in a fluid flow satisfies the Stokesian flow assumption, and the settling speed is expressed as \cite{Kim:2005}:
\begin{equation}
	\label{eqVsettle}
	\ve{v}_{\rm settle} = -v_1 \ve{e}_y 
	- \left(v_3-v_1 \right)\left(\ve{e}_y\cdot\ve{n} \right) \ve{n},
\end{equation}
where $v_1$ and $v_3$ are the Stokesian terminal velocities of a spheroid in a quiescent fluid with its symmetry axis orientated orthogonal to and parallel to the gravity direction, respectively. 
We specify the direction of $y$-axis $\ve{e}_y$ as the direction opposite to gravity, i.e. $\ve{e}_y = -\ve{g}/\left| \ve{g}\right|$. 

The swimmer's angular velocity is expressed as \cite{Gustavsson2019}:
\begin{equation}
	\label{eqFIswimmer}
		\ve{\omega}_{\rm p} = \frac{1}{2}\ve{\omega} 
		+ \frac{\lambda^2-1}{\lambda^2+1}\left(\ve{n}\wedge \mat S\cdot\ve{n} \right) 
		+ \frac{M}{\gamma}\left[v_{\rm swim}v_1\left(\ve{e}_y\wedge\ve{n} \right)
		- v_1v_3\left(\ve{e}_y\cdot\ve{n} \right)\left(\ve{e}_y\wedge\ve{n} \right) \right],
\end{equation}
where the first two terms on the right-hand-side originate from the Jeffery torque \cite{Jeffery1922}, which represent the contributions of local fluid vorticity $\boldsymbol{\omega} $ and strain rate $\mat{S}$, respectively.
The aspect ratio $\lambda$ is defined as the ratio of the lengths between the major and minor axes of the spheroidal swimmer, with $\lambda=1$ for spheres and $\lambda>1$ for elongated spheroids. 
The third term on the right-hand-side of Eq.~(\ref{eqFIswimmer}) is the contribution of fluid inertial torque, where the shape factor $M$ only depends on $\lambda$. $M$ is zero for spheres and negative for elongated spheroids, ranging from $M=0$ for $\lambda=1$, to $M\approx -0.1$ when $\lambda$ ranges from 2 to 8 (see Appendix A). Therefore, spherical swimmers are not subjected to fluid inertial torque.
The contribution of fluid inertial torque consists of two parts. For convenience, we call the term $M v_{\rm swim}v_1\left(\ve{e}_y \wedge \ve{n} \right)/\gamma$ \emph{swimming-settling term}, which denotes the coupling effect of swimming and settling. Similarly, we call $-Mv_1v_3\left(\ve{e}_y\cdot\ve{n} \right)\left(\ve{e}_y \wedge \ve{n} \right)/\gamma$ the \emph{settling term}, which is only ascribed to settling effect.

\begin{figure}
	\centering
	\includegraphics[width=7.8cm]{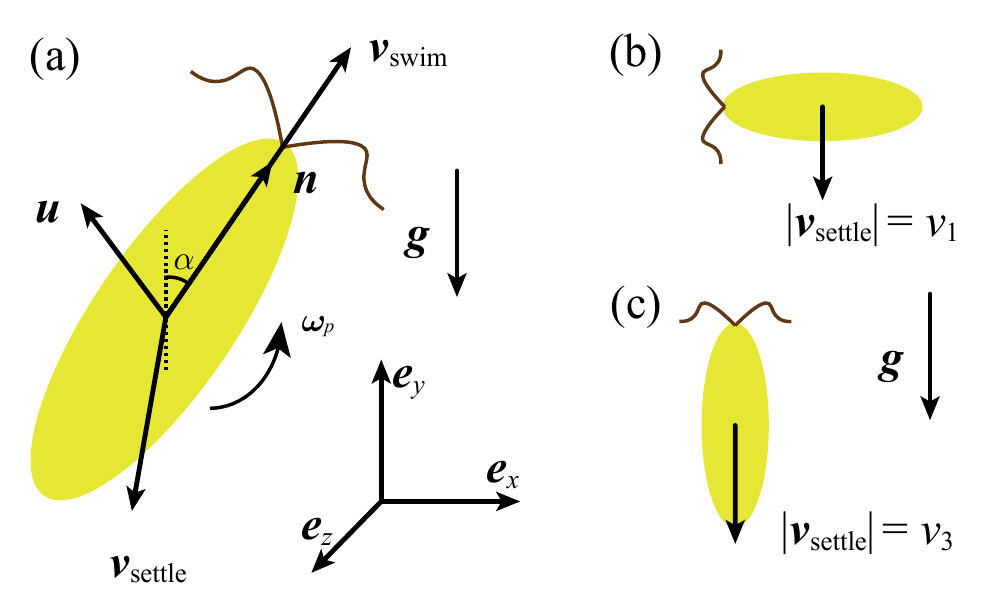}
	\caption{(a) A sketch of a settling elongated swimmer. $\ve{e}_x$, $\ve{e}_y$, and $\ve{e}_z$ are the base vectors of the global frame of reference. (b) A swimmer settling with the symmetry axis perpendicular to gravity. (c) A swimmer settling with its symmetry axis parallel to the direction of gravity. }
	\label{figsketch}
\end{figure}


\section*{Swimmers in a quiescent fluid}
To understand how fluid inertial torque affects the orientation of swimmers, we first analyze angular dynamics in a quiescent fluid.
Using Eq.~(\ref{eqFIswimmer}), the rotation of a swimmer is described as:
\begin{equation}
	\frac{\dif\alpha}{\dif t} = \frac{M}{\gamma}\left( v_{\rm swim}v_1 \sin{\alpha} - v_1v_3\cos{\alpha} \sin{\alpha}\right).
	\label{eqalpha}
\end{equation}
Here, $\alpha$ is the angle of $\ve{n}$ relative to $\mathbf{e}_y$ [Figure~\ref{figsketch}(a)], and thus $n_y \equiv \ve n\cdot\ve e_y = \cos\alpha$.
From Eq.~(\ref{eqalpha}), a swimmer has three equilibrium orientations:
\begin{equation}
\alpha_0^{(1)} = 0,~ \alpha_0^{(2)}=\arccos{\frac{v_{\rm swim}}{{v_3}}},~\text{and}~ \alpha_0^{(3)}=\pi,
\label{eqEqui}
\end{equation}
which correspond to (1) swimming upward against gravity, (2) swimming with a fixed angle relative to gravity direction, and (3) swimming downward along the gravity direction, respectively. Derived from Eq.~(\ref{eqalpha}), the first order linear equation of a small perturbation around the equilibrium orientations, $\delta_\alpha$, reads:
\begin{equation}
\frac{\dif{\delta_\alpha}}{\dif t} = 
\frac{Mv_1v_3}{\gamma} \left(R_v\cos{\alpha_0}-\cos{2\alpha_0} \right)\delta_\alpha,
\label{eqPert}
\end{equation}
where $R_v=v_{\rm swim}/v_3$. The solution of Eq.~(\ref{eqPert}) is 
\begin{equation}
	\delta_\alpha = 
	\exp\left[\frac{Mv_1v_3}{\gamma} \left(R_v\cos{\alpha_0}-\cos{2\alpha_0} \right) t \right].
	\label{eqPert2}
\end{equation}
Inserting Eq.~(\ref{eqEqui}) into (\ref{eqPert2}), and with $M<0$ we find $\alpha_0^{(3)}$ is always unstable, and there is one and only one stable orientation between $\alpha_0^{(1)}$ and $\alpha_0^{(2)}$, depending on the value of $R_v$. 
In general, the stable orientation is $n_{y,0} = \min\left(1,R_v \right)$.

\begin{figure}
	\centering
	\includegraphics[width=8cm]{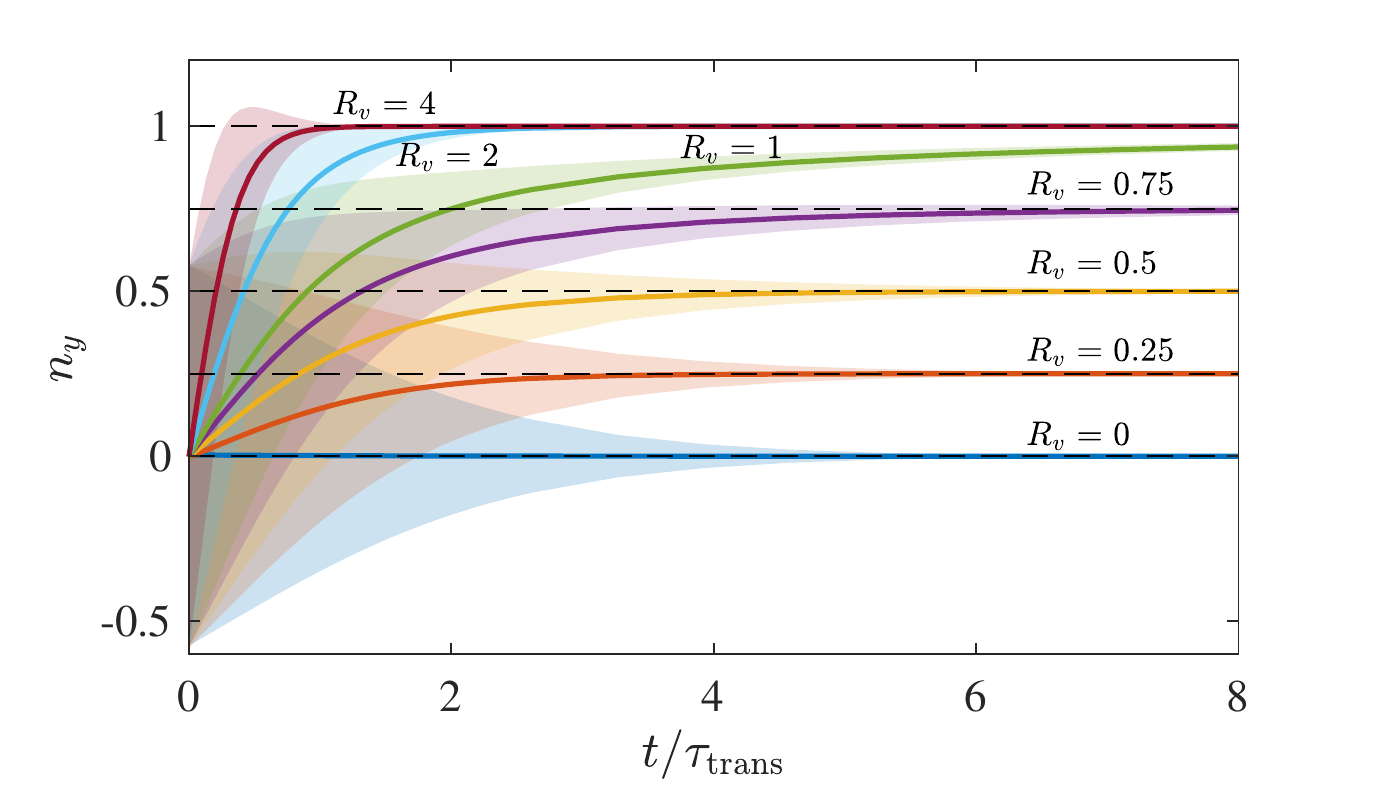}
	\caption{Evolution of the orientation of swimmers $n_y$ over dimensionless time $t/\tau_{\rm trans}$ in a quiescent fluid, where $\tau_{\rm trans} = \gamma / \left|M \right|v_1v_3$ is a time scale for the transient regime according to Eq.~(\ref{eqPert2}). Solid lines represent the mean value of $n_y$, and the colored areas represent the ranges of mean $\pm$ standard deviation. Horizontal dashed lines stand for the theoretical equilibrium orientation. Aspect ratio is $\lambda = 8$ for all cases here. }
	\label{figqf}
\end{figure}

The dependence of $n_{y,0}$ on $R_v$ is due to the competition between the swimming-settling term and the settling term, which make opposite contributions to the orientation. The swimming-settling term tends to align a swimmer in the upward direction, whereas the settling term tends to align the swimmer horizontally as indicated in Refs.~\cite{Sheikh2020,Gustavsson2019}.
When $R_v<1$, the swimming-settling term does not overcome the settling term, so the swimmer reaches an inclined orientation where the two terms are balanced. 
When $R_v\ge1$, the swimming-settling term overcomes the settling term for any orientation, so the swimmer rotates to swim upward. Simulations in a quiescent fluid are performed to verify the aforementioned theoretical analysis. Figure~\ref{figqf} shows that swimmers with random initial orientation gradually approach the theoretical equilibrium orientation after a transient time. In the critical case of $R_v = 1$, swimmers takes much longer time to approach the stable orientation, because swimming-settling term and the settling term are almost balanced at $n_y\approx 1$, resulting in a small angular velocity.

\section{Effective reorientation due to fluid inertial torque}
Swimmers with $R_v \ge 1$ spontaneously swim in upward direction, which are similar to the well-known gyrotactic swimmers with bottom-heaviness \cite{Kessler1986} or fore-aft asymmetry \cite{O'Malley2011,Roberts1970}. 
The similarity can also be deduced from Eq.~(\ref{eqFIswimmer}).  The fluid inertial term in Eq.~(\ref{eqFIswimmer}) for elongated swimmers can be written as $-(\ve e_y \times \ve n)/2B_I$, where 
\begin{equation}
B_I = \frac{\gamma}{
		2|M|v_{\rm swim}v_1}\left[1 -\frac{v_3}{v_{\rm swim}} \left(\ve{e}_y\cdot\ve{n} \right)\right]^{-1}.
\label{eqBI}
\end{equation}
This is similar to the widely used model of regular gyrotaxis $-\left(\ve{e}_y\times\ve{n} \right)/2B$~ \cite{pedley1987orientation,Durham2013,Gustavsson2016}, where $B$ is the reorientation time scale which quantifies how fast a swimmer recovers its stable orientation under gyrotactic torque.
$B_I$ can be regarded as an effective reorientation time scale provided by fluid inertial torque if $B_I>0$.

Eq.~(\ref{eqBI}) shows some characteristics of $B_I$. First, only non-spherical, settling swimmers experience the effective gyrotaxis. Spherical or non-settling swimmers have a zero $M$ or $v_1$, which yields infinite $B_I$ (zero fluid inertial torque). Second, $B_I$ depends on the instantaneous orientation of a swimmer because of the contribution of settling term. The dependence on orientation complicates the problem, because reorientation time scale varies along the trajectory as the swimmer rotates, and posterior knowledge of the mean orientation of swimmers is required to estimate the magnitude of fluid inertial torque.  However, $B_I$ is almost constant if $\langle\ve e_y \cdot \ve n \rangle v_3/v_{\rm swim} \approx 0$, i.e. the settling term is negligible.
This is justified when a swimmer swims much faster than it settles, or when a swimmer has $\langle\ve e_y \cdot \ve n \rangle\approx 0$ along its trajectory. The first condition is true for typical plankton species (see Tables \ref{table1} and \ref{table2} in Appendix C), and the latter is true when fluid inertia torque is weak and the rotation of a swimmer is dominated by random turbulent fluctuations.
When we neglect the setting term, $B_I$ is expressed as
\begin{equation}
B_I \approx \frac{\gamma}{
	2|M|v_{\rm swim}v_1}.
\label{eqBIsimplify}
\end{equation}
With Eq.~(\ref{eqBIsimplify}) we can quantify the magnitude of fluid inertial gyrotaxis. When a swimmer settles or swims faster, or when it has a larger $M$, it experiences a stronger gyrotactic effect caused by fluid inertial torque.

\section*{Orientation of swimmers in turbulence}

\begin{figure}
	\includegraphics[width=8.6cm]{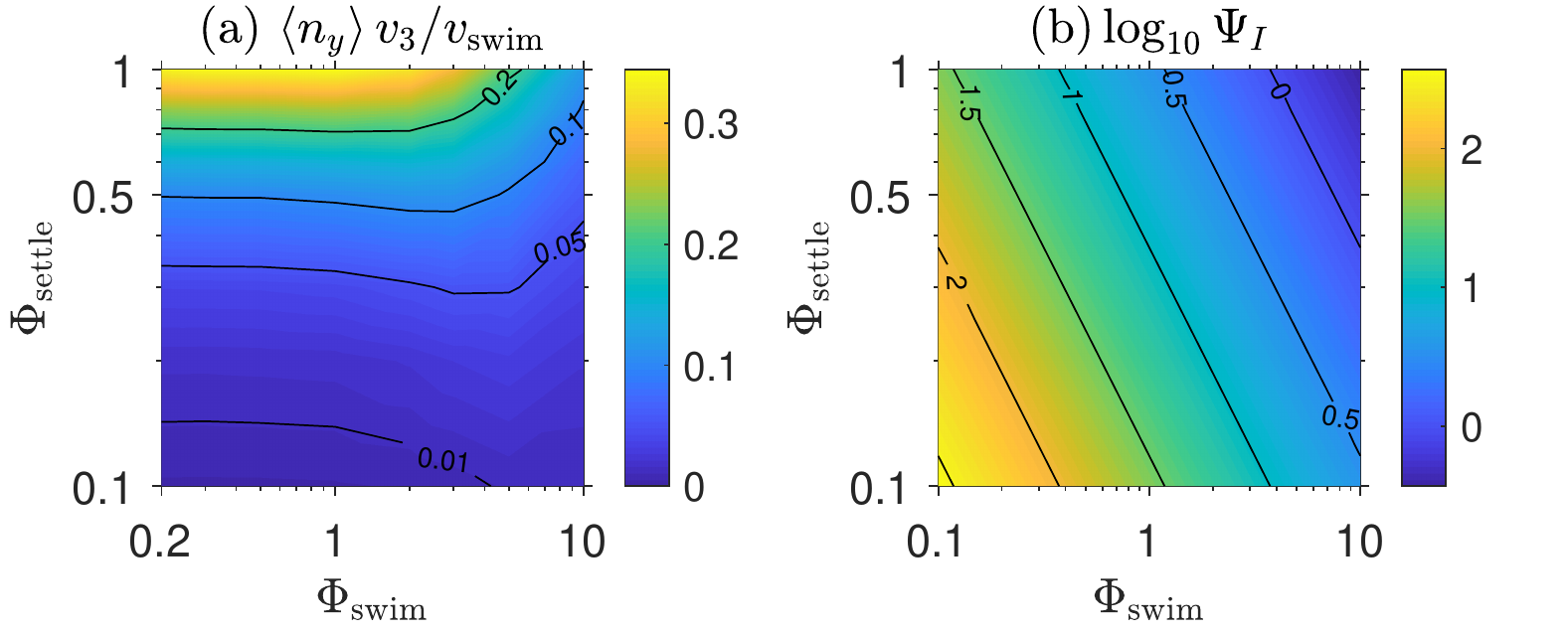}
	\caption{(a) Value of $\left\langle n_y \right\rangle v_3 /v_{\rm swim}$ in the parameter range of $\Phi_{\rm swim}$ and $\Phi_{\rm settle}$, using $\left\langle n_y \right\rangle$ of swimmers with $\lambda=8$ as example. (b) Value of $\Psi_I$ calculated with Eq.~(\ref{eqBIsimplify}) in the parameter range of $\Phi_{\rm swim}$ and $\Phi_{\rm settle}$ in the present study, with $M=-0.135$ corresponding to swimmers with $\lambda=8$.}
	\label{figPsiI}	
\end{figure}

\begin{figure}
	\centering
	\includegraphics[width=8.6cm]{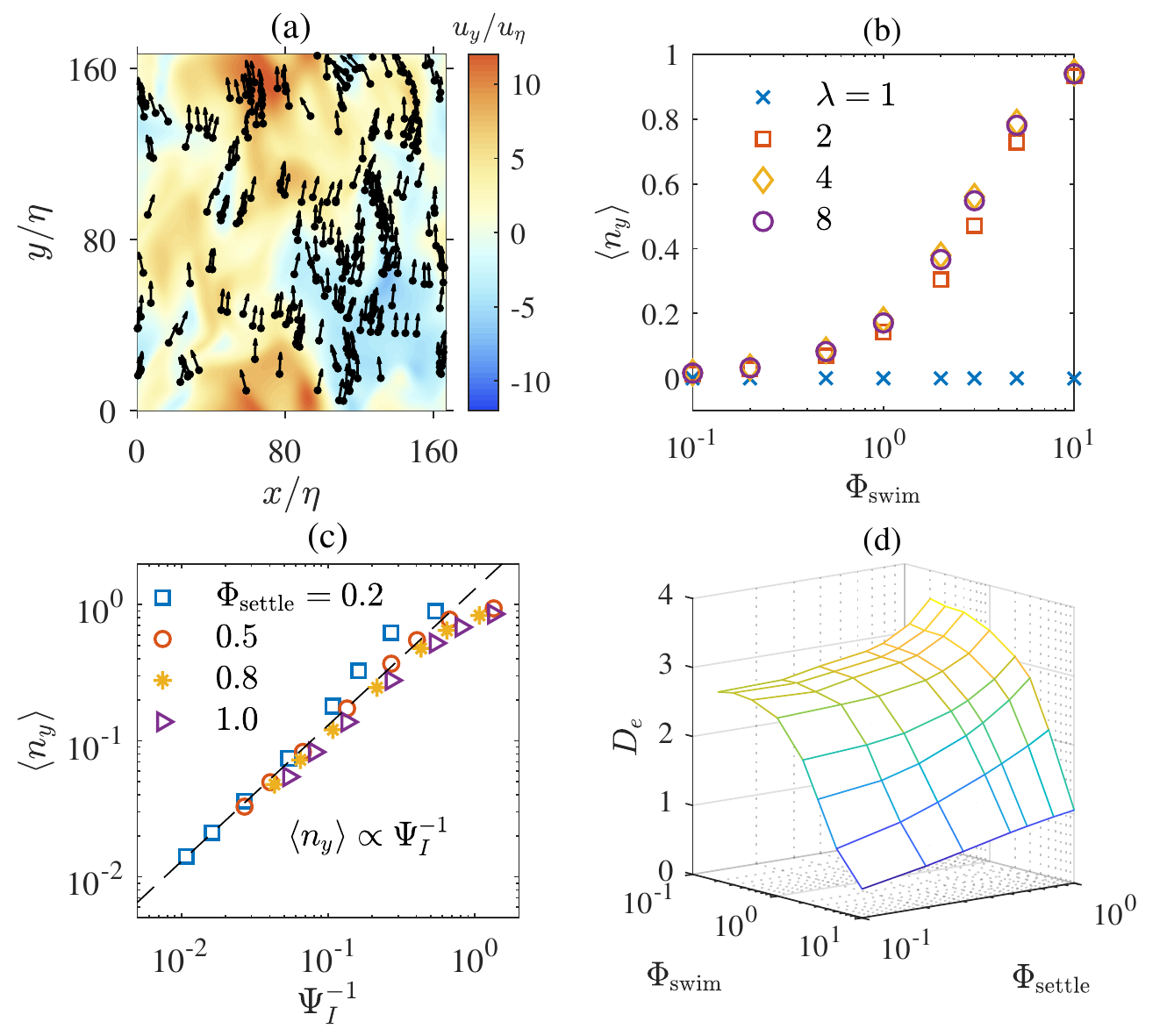}
	\caption{(a) Instantaneous spatial distribution of swimmers in HIT. Black dots and tiny arrows stand for the position and swimming direction of each swimmer, respectively. Colors represent the vertical fluid velocity $u_y$. Quantities are normalized by Kolmogorov velocity and length scale, $u_{\eta}$ and $\eta$, respectively. Parameters of swimmers are $\phisettle= 0.5$, $\phiswim=10$, and $\lambda = 2$, corresponding to $\Psi_I=0.99$.
	(b)~$\left\langle n_y\right\rangle$ of swimmers with different aspect ratio and $\phisettle=0.5$ in HIT.  
	(c)~Mean orientation $\left\langle n_y\right\rangle$ versus $\Psi_I^{-1}$ with $\lambda$ = 8. The slope of the dashed line represent the relationship of $\left\langle n_y\right\rangle \propto \Psi_I^{-1}$.
	(d)~$D_e$ as a function of $\phiswim$ and $\phisettle$, obtained by fitting distribution~(\ref{eqdistrib}).
			}
	\label{figalign-phi}
\end{figure}

Planktonic micro-swimmers in the ocean or estuaries live in turbulent environment, and their orientation controls the direction and efficiency of vertical migration. Therefore, it is necessary to understand how fluid inertial torque influences the orientation of swimmers in turbulence.
We use Eulerian-Lagrangian direct numerical simulations to obtain the trajectories of swimmers in a forced homogeneous isotropic turbulence (HIT), and focus on the statistics of orientation. 
The HIT has a Taylor Reynolds number $\mathrm{Re}_\lambda = u_{\rm rms}^2 \sqrt{15/(\gamma\epsilon)} = 60$, where $u_{\rm rms}$ and $\epsilon$ are the root-mean square velocity and dissipation rate, respectively.
The incompressible Navier-Stokes equations are solved by a pseudo-spectral method with $96^3$ grid points to ensure the accuracy of resolution at small scales.
Statistics of each parameter configuration are obtained by averaging over 40 uncorrelated time samples of 120,000 trajectories. Details of numerical methods are provided in Appendix B.

First, we need to quantify the magnitude of fluid inertial gyrotaxis relative to the turbulent motion.
Turbulence influences the rotation of swimmers by the fluid velocity gradients along their trajectories as shown in Eq.~(\ref{eqFIswimmer}).
The magnitude of the velocity gradients in a turbulent flow is of the order of Kolmogorov time scale $\tau_{\eta}$.  
Thus, we normalize $B_I$ with $\tau_{\eta}$ and obtain
\begin{equation}
\Psi_I \equiv B_I/\tau_{\eta} \approx (2|M|\Phi_{\rm swim}\Phi_{\rm settle})^{-1},
\label{eqpsiI}
\end{equation}
where $\phiswim = v_{\rm swim}/u_{\eta}$ and $\phisettle = \left(2v_1+v_3\right)/3u_{\eta}$ are the dimensionless swimming and settling speeds.
Note that we assume $v_1\approx v_3$ and thus $\phisettle \approx v_1 /u_{\eta}$ in Eq.~(\ref{eqpsiI}) based on the fact that $1 \le v_3/v_1 < 1.7$ for spheroids (see Appendix A). Similar to the parameter $\Psi$ for gyrotaxis widely used in Refs.~\cite{Durham2013,Gustavsson2016,Zhan2014,Borgnino2018}, 
$\Psi_I$ quantifies the effective gyrotaxis provided by fluid inertial torque.
According to the typical values for oceanic plankton~\cite{Lovecchio2019,Sengupta2017,Smayda2010,Titelman2003,Kamykowski1992} (and also see Appendix C), we investigate swimmers within a parameter space of $0\le\phiswim\le~10$, $0\le\phisettle \le 1$, and $1\le\lambda \le 8$. In most of this parameter range, the settling term can be neglected because $\left\langle n_y \right\rangle v_3 /v_{\rm swim} \approx 0$ as shown in Figure~\ref{figPsiI}(a). Accordingly, we calculate the range of $\Psi_I$ using Eq.~(\ref{eqpsiI}). Figure~\ref{figPsiI}(b) shows that $\Psi_I$ varies over two orders of magnitude in the present study. The decrease of $\Psi_I$ at increasing $\phiswim$ and $\phisettle$ reflects that the ratio between the contributions of fluid inertial torque and Jeffery’s torque increases as the relative velocity between the swimmer and the fluid grows.

Figure \ref{figalign-phi}(a) shows the instantaneous spatial distribution and orientation of swimmers with $\Psi_I = 0.99$, $\lambda =8$. We observe an obvious preferential alignment in the upward direction, because the swimmers are subjected to a fluid inertial torque of the order of fluid velocity gradients.
However, only elongated swimmers obtain upward orientation as shown in Figure~\ref{figalign-phi}(b). Eqs.~(\ref{eqBI})-(\ref{eqpsiI}) show that the reorientation time is proportional to $|M|^{-1}$. Spherical swimmers have $M=0$, which indicates the fluid inertial torque vanishes and the reorientation time approaches infinity. In this case, the orientation of swimmers is almost isotropic, and $\langle n_y \rangle = 0$.
On the contrary, elongated swimmers preferentially align in the upward direction. $|M|$ is non-monotonous to $\lambda$, which reaches the maximum at about $\lambda=4$ (see Appendix A). Therefore, among the four aspect ratios we considered, $\left\langle n_y\right\rangle$ is the largest when $\lambda = 4$ (Figure~\ref{figalign-phi}(b)), in which case $\Psi_I$ is minimal.
We note that swimmers with $\lambda =2$ already show a strong preferential orientation in the upward direction, which means fluid inertial torque can be significant even for slightly elongated swimmers.
	
Figure~\ref{figalign-phi}(c) shows the relation between $\Psi_I$ and the orientation of swimmers. We observe that $\left\langle n_y\right\rangle$ is approximately proportional to $\Psi_I^{-1}$, suggesting a strong correlation between the orientation of swimmers and $\Psi_I$. 
The linearity is the best when $\Psi_I$ is large, which can be explained by the probability distribution of orientation of swimmers. For weak gyrotactic swimmers, the fluctuating turbulent velocity gradients act as Gaussian noises, and the rotation of gyrotactic swimmers are diffusive \cite{fouxon2015phytoplankton}. In this case, the orientation of spherical gyrotactic swimmers obeys an equilibrium distribution \cite{lewis2003orientation,fouxon2015phytoplankton}:
\begin{equation}
g(n_y) = \frac{\beta e^{n_y/\beta}}{2\sinh\beta}, ~\text{with}~ \beta = \frac{1}{\Psi \tau_\eta D_{e}},
\label{eqdistrib}
\end{equation}
where $\Psi$ is the gyrotactic parameter, and the effective rotation diffusivity $D_e$ is determined by the time correlation of velocity gradients along the trajectories of swimmers, i.e. $D_e \sim \tau_{cor} / \tau_\eta^2$~\cite{fouxon2015phytoplankton}, where $\tau_{cor}$ is the correlation time.
Eq.~(\ref{eqdistrib}) is derived for spherical, non-settling swimmers, but we have verified that Eq.~(\ref{eqdistrib}) fits well with the distribution of settling elongated swimmers under fluid inertial torque in present study. Figure~\ref{figalign-phi}(d) shows the best-fit $D_e$ for different $\phiswim$ and $\phisettle$.
$D_e$ is expected to have little dependence on $\phiswim$ and $\phisettle$ when they are both much smaller than unity (which is the case for large $\Psi_I$). In this case, swimmers have small relative velocity with respect to the fluid, and they almost follow streamlines. Therefore, the correlation time scale of fluid velocity gradients along their trajectories is $\tau_{cor} \sim \tau_\eta$, so that $D_e \sim \tau_\eta^{-1}$~\cite{fouxon2015phytoplankton} and that $\beta \sim \Psi^{-1}$. 
Moreover, the probability distribution~(\ref{eqdistrib}) gives the mean orientation $\langle n_y \rangle = \coth\beta - \beta^{-1}$, which yields $\langle n_y \rangle \propto \beta$ for small $\beta$.
This gives $\langle n_y \rangle \propto \Psi_I^{-1}$ for large $\Psi_I$ as shown in Figure~\ref{figalign-phi}(c).

\section*{Conclusions}
The present study investigates the significance of fluid inertial torque on settling micro-swimmers owing to the velocity difference between the swimmers and fluid. The effect of fluid inertial torque shares a similar mathematical form with regular gyrotaxis mechanisms caused by bottom-heaviness or fore-aft asymmetry. The fluid inertial torque stabilizes the orientation of swimmers and allows them to swim in the upward direction spontaneously.
Therefore, we suggest that fluid inertial torque is an effective mechanism of gyrotaxis for elongated settling swimmers.

The magnitude of fluid inertial torque depends on the shape, swimming and settling speeds of a swimmer. Similar to Ref.~\cite{pedley1987orientation}, we quantify the gyrotactic effect produced by fluid inertial torque by $B_I$, which is an effective reorientation time measuring how fast a swimmer restores its stable orientation under fluid inertial torque. From $B_I$ we know some characteristics of fluid inertial gyrotaxis. 
First, only elongated, settling swimmers are subject to fluid inertial torque because they have non-zero shape factor $M$ and settling velocity $v_1$ and $v_3$. 
Second, fluid inertial torque is stronger when swimmers swim and settle faster, in which case $B_I$ is small. 
Third, $B_I$ depends on the instantaneous orientation of swimmers due to the effect of settling term in Eq.~(\ref{eqFIswimmer}), but in the limit of $v_{\rm swim} \gg v_3$, $B_I$ is nearly independent with orientation and can be approximated by Eq.~(\ref{eqBIsimplify}). This limit holds for typical plankton species, and it allows for predicting $B_I$ from the gaits of plankton without knowing their real-time orientation.

The orientation of swimmers under fluid inertial torque in turbulence is strongly related to the dimensionless parameter $\Psi_I$. When $\Psi_I \le 1$, swimmers show strong alignment with upward direction, yielding  $\langle n_y \rangle \approx 1$. When $\Psi_I \gg 1$, $\langle n_y \rangle \propto \Psi_I^{-1}$ as a result of the diffusive effect of turbulent fluid velocity gradients. We also show that swimmers with $\lambda = 2$ is strongly affected by fluid inertial torque, which implies that fluid inertial torque can be significant even when swimmers are not strongly elongated.

Fluid inertial torque may have a potential impact on the gyrotaxis for elongated planktonic swimmers, especially for those forming long chains and thus having large swimming and settling speeds \cite{Smayda2010,Davey1985}. Settling effect of micro-swimmers was often neglected in previous studies~\cite{Lovecchio2019,Durham2009,Sengupta2017,Durham2013,DeLillo2014,Zhan2014,Gustavsson2016,Borgnino2018}. However, our results demonstrate that neglecting settling will lead to an underestimation of gyrotaxis because fluid inertial torque vanishes without settling effect.
Moreover, unlike the two well-known gyrotaxis mechanisms which contribute to the rotation dynamics passively, the fluid inertial torque can be tuned by the swimming speed. This feature provides possibility for micro-swimmers to actively control the gyrotactic reorientation time by adjusting their swimming velocity.
As a new mechanism of gyrotaxis, swimmers under fluid inertial torque are also expected to sample specific flow regions and form local clustering as bottom-heavy gyrotactic swimmers do~\cite{Durham2013,Borgnino2018,Zhan2014}. These phenomena are known to be controlled by the dimensionless reorientation time $\Psi$ and the swimming speed $\phiswim$. In the case with fluid inertial torque, one has to consider the influence of settling as well, because it influence the reorientation time $\Psi_I$.

The present study focuses on the fluid inertial torque induced by the relative velocity between swimmer and local fluid. We note that the current swimmer model is still idealized. For instance, it neglects the influence of fluid velocity gradients and unsteadiness on the fluid inertial torque\cite{candelier2016angular,candelier2019time}. These effects could be important for swimmers in flows with strong shear, and deserve to be studied in the future.

%
%
%

\begin{acknowledgments}
	This work was supported by the National Natural Science Foundation of China (Grant No. 11911530141 and 91752205). JQ and LZ acknowledge the support from the Institute for Guo Qiang of Tsinghua University (Grant No. 2019GQG1012).
\end{acknowledgments}

\section*{Appendix A. Inertia-less point-particle model for a settling swimmer}
Following Refs. \cite{Gustavsson2019,Sheikh2020}, here we derive the governing equations for an elongated settling micro-swimmer (Eqs.~(1) to (4) in the paper). The Newton's second law for a spheroidal particle writes:
\begin{equation}
	\massp \frac{\mathrm{d}\vp}{\mathrm{d}t} = \ve F, \label{newton1}
\end{equation}
\begin{equation}
	\massp \frac{\mathrm{d}}{\mathrm{d}t}\left[\mat{I}_p(\ve n)\ve \omega_p\right] = \ve T.
\end{equation}
Here, $\massp$ is the particle mass, $m_{\rm f}$ is the mass of fluid occupied by the particle, $\ve v_p$ and $\ve \omega_p$ are the particle velocity and angular velocity, respectively. $\mat{I}_p = I_{p,ij}$ is the rotational inertia tensor per unit-mass of the particle,
\begin{equation}
	I_{p,ij} = I_{\perp}(\delta_{ij}-n_i n_j)+I_{\parallel}n_i n_j,
\end{equation}
where $\ve n$ is the particle swimming direction, $I_{\perp} = a^2(1+\lambda^2)/5 $ and $I_{\parallel} = 2a^2/5$, with $a$ being the half length of the minor axis of the particle, and $\lambda$ is the aspect ratio defined as the ratio between the length of the major and minor axes of the particle. The force on a swimmer reads:
\begin{equation}	
	\ve F =\ve F_{St} + \ve F_{I} + F_{swim}\ve n + \left(\massp - m_{\rm f}\right) \ve g,	
\end{equation}
\begin{equation}
	\ve F_{St} =6\pi a \rho_{\mathrm f} \gamma \mat{A}(\ve u - \ve v_p), \label{FST} 
\end{equation}
\begin{equation}
	\ve F_{I} = \frac{9\pi}{8} \rho_{\mathrm{f}}a^2\max(\lambda,1)\left|\vp - \ve u\right|  \\  
	\times \left[3\mat A - \mat I(\hat{\ve{u}}_s \cdot \mat A \hat{\ve{u}}_s)\right] \mat A(\ve u - \vp). \label{FI}
\end{equation}
The total force is the summation of Stokes drag force $\ve F_{St}$ \cite{brenner1963stokes}, the fluid inertial correction of force, or so-called Oseen correction, $\ve F_{I}$ \cite{brenner1961oseen,khayat1989inertia}, the swimming propulsion force $F_{swim}\ve n$, and the contributions of gravity and buoyancy $\left(\massp -m_{\rm f}\right)\ve g$. 
In Eq.~(\ref{FST}), $\rho_{\mathrm f}$ and $\gamma$ are the density and kinematic viscosity of fluid, respectively, and $\ve u$ is the fluid velocity at the particle position.
The translational resistant tensor $\mat A$ is defined as:
\begin{equation}
	A_{ij} = A_{\perp}(\delta_{ij} - n_i n_j) + A_{\parallel}n_i n_j,
\end{equation}
where $A_{\perp}$ and $A_{\parallel}$ depend only on the aspect ratio $\lambda$ of a particle, and the expressions can be found in Refs. \cite{Gustavsson2019,kim2013microhydrodynamics}. In Eq.~(\ref{FI}), $\hat{\ve u}_s = (\ve u - \vp)/\left|\ve u - \vp\right|$.

The torque on a particle reads:
\begin{align}
	\ve T =~& \ve T_{J} + \ve T_{I}, \\
	\ve T_{J} =~& 6\pi a\rho_{\mathrm f}\gamma\mat C \left(\frac{1}{2}\ve \omega - \ve\omega_\mathrm{p} \right) +  6\pi a\rho_{\mathrm f}\gamma\mat H:\mat S,\\
	\ve T_{I} =~& F_{\beta}\rho_\mathrm{f} a^3\max(\lambda,1)^3
	\times \left[\ve n \cdot \left(\vp - \ve u\right) \right] \left[\ve n \wedge \left(\vp - \ve u\right) \right].\label{TI}
\end{align}
Here, the Jeffery torque $\ve T_{J}$ is related to the local fluid vorticity $\ve \omega$ and strain rate $\mat S$ \cite{Jeffery1922}, and $\mat C$ and $\mat H$ are rotational resistant tensors \cite{kim2013microhydrodynamics,Gustavsson2019}:
\begin{gather}
	C_{ij}  = C_{\perp}(\delta_{ij}-n_i n_j) + C_{\parallel} n_i n_j,\\
	H_{ijk}= H_0 \epsilon_{ijl} n_k n_l,
\end{gather}
where $ C_{\perp}$, $C_{\parallel}$ and $H_0$ are given in Refs. \cite{Gustavsson2019,kim2013microhydrodynamics}. 
The fluid inertial torque $\ve T_{I}$ depends on both the magnitude and direction of the relative translational motion between particle and fluid \cite{Dabade2015}, and thus, influences the orientation of particles whenever they translate relative to the fluid, such as settling \cite{Gustavsson2019,Sheikh2020} or swimming.
In Eq.~(\ref{TI}), $F_\beta$ is a parameter depending only on $\lambda$ \cite{Dabade2015}.

Now, we can evaluate the relative importance of fluid inertial force and torque by comparing their magnitudes with those of other terms. For inertial force, $\left|\ve F_{I}\right| / \left|\ve F_{St}\right| \sim  a \left| \ve u- \ve v_{\mathrm p}\right| / \gamma = \Rep$. This suggests that $\ve F_{I}$ is negligible in the limit of $\Rep \ll 1$. However, as discussed in Ref. \cite{Sheikh2020}, the magnitudes of fluid inertial and Jeffery torques scale differently:
\begin{equation}
	\frac{\left|\ve T_{I}\right|}{\left|\ve T_{J}\right|} \sim \frac{\left|\ve u-\vp\right|^2}{\left|\ve \omega/2 - \ve \omega_{\rm p}\right|\gamma},	
\end{equation}
suggesting that the fluid inertial torque cannot be neglected even in the limit of $\Rep \ll 1$. Therefore, in the following derivation, we neglect the inertial correction for the drag force but still keep the inertial torque.

Using Eqs.~(\ref{newton1}) to (\ref{TI}), we obtain the governing equations of particle motion \cite{Gustavsson2019}:
\begin{equation}	
	\frac{\mathrm{d}{\vp^\prime}}{\mathrm{d}t} = \frac{1}{\rm{St}}\left[\mat A \left(\ve u^\prime -\vp^\prime\right)
	+ \frac{F_{swim}\tau_{\rm p}}{\massp u_\eta} \ve n \right.
	+\left. \frac{\left(\rho_{\rm p}-\rho_{\rm f}\right)g\tau_p }{\rho_{\rm p}u_\eta} \ve e_g
	\right],\label{vpdot}		
\end{equation}
\begin{equation}
		\frac{\mathrm{d}{\ve \omega_{\rm p}^\prime}}{\mathrm{d}t} =  \frac{1}{\rm{St}} 
		\bigg[
		\mat I_p^{-1}\mat C^\prime\left(\frac{1}{2}\ve \omega^\prime-\ve\omega_{\rm p}^\prime\right) 
		+ \mat I_p^{-1} \mat H^\prime:\mat S^\prime 
		+ \mathcal{A} \left[\ve n \cdot \left(\vp^\prime - \ve u^\prime\right) \right] \left[\ve n \wedge \left(\vp^\prime - \ve u^\prime\right) \right] \bigg] 
		+ \frac{\lambda^2-1}{\lambda^2+1} \left(\ve n \cdot \ve \omega_{\rm p}^\prime\right) \left(\ve \omega_{\rm p}^\prime \wedge \ve n \right). 
		\label{omegapdot}
\end{equation}

Here, quantities with primes are non-dimensionalized by Kolmogorov velocity scale $u_\eta$ and time scale $\tau_\eta$. We note that the inertial force correction $\ve F_{I}$ has been neglected for the derivation of Eq.~(\ref{vpdot}) with $\Rep \ll 1$, and we use the relationship $\mathrm{d}(\mat{I}_p\ve\omega_p)/\mathrm{d}t = \mat{I}_p \mathrm{d}\ve \omega_p/\mathrm{d}t + \ve\omega_p \wedge (\mat{I}_p\ve\omega_p)$ in Eq.~(\ref{omegapdot}).
The Stokes number ${\rm St}=\tau_{\rm p} / \tau_{\eta}$ quantifies the inertia of the swimmer, where $\tau_{\rm p} \equiv (2a^2\lambda \rho_{\rm p})/(9\gamma\rho_{\rm f})$ is the particle translational response time, and $\tau_{\eta}$ is the Kolmogorov time scale. For typical plankton species, $\rm{St}$ is usually much smaller than unity as shown in Table~\ref{table2}.
In the limit of $\rm{St} \ll 1$, i.e. overdamped limit~\cite{Gustavsson2019}, Eqs.~(\ref{vpdot}) and (\ref{omegapdot}) can be further simplified:
\begin{subequations}
	\label{eom}
	\begin{equation}
		\ve{v}_{\rm p} = \ve{u} + v_{\rm swim} \ve{n} + \ve{v}_{\rm settle},
	\end{equation}
	\begin{equation}
		\label{eqFI}
		\ve{\omega}_{\rm p} = \frac{1}{2} \ve{\omega} 
		+ \Lambda\left(\ve{n}\wedge\mat{S}\cdot\ve{n} \right) 
		+ \frac{M}{\gamma}\left[\ve{n}\cdot\left(\ve{v}_{\rm p}-\ve{u} \right) \right]
		\left[\ve{n}\wedge\left(\ve{v}_{\rm p}-\ve{u} \right) \right],
	\end{equation}
\begin{equation}
		\text{where}~~ v_{\rm swim}\ve n = \frac{F_{swim}\tau_{\rm p}}{\massp} A_{\parallel}^{-1} \ve n,
\end{equation}		
\begin{equation}	
		\text{and}~~ \ve v_{\rm settle} = \frac{\rho_{\rm p}-\rho_{\rm f}}{\rho_{\rm p}}g\tau_{\rm p}\mat A^{-1} \ve e_g.	
\end{equation}
In the present study, we directly assign the value of $v_{\rm swim}$, and we use an equivalent definition of $\ve v_{\rm settle}$ \cite{Kim:2005}:
\begin{equation}	
	\ve{v}_{\rm settle} = -v_1 \ve{e}_y 
	- \left(v_3-v_1 \right)\left(\ve{e}_y\cdot\ve{n} \right) \ve{n},\\
\end{equation}
\begin{align}
	\text{with}~~\ve e_y &= -\ve e_g,\nonumber\\
	v_1 &= \frac{\rho_{\rm p}-\rho_{\rm f}}{\rho_{\rm p}}g\tau_{\rm p}A_{\perp}^{-1},\nonumber\\
	v_3 &= \frac{\rho_{\rm p}-\rho_{\rm f}}{\rho_{\rm p}}g\tau_{\rm p}A_{\parallel}^{-1},\nonumber
\end{align}
\end{subequations}
where $v_1$ and $v_3$ are the terminal settling speeds of a spheroid in quiescent fluid, with symmetry axis perpendicular and parallel to gravity direction, respectively.

\begin{figure}
	\centering
	\includegraphics[width=7.0cm]{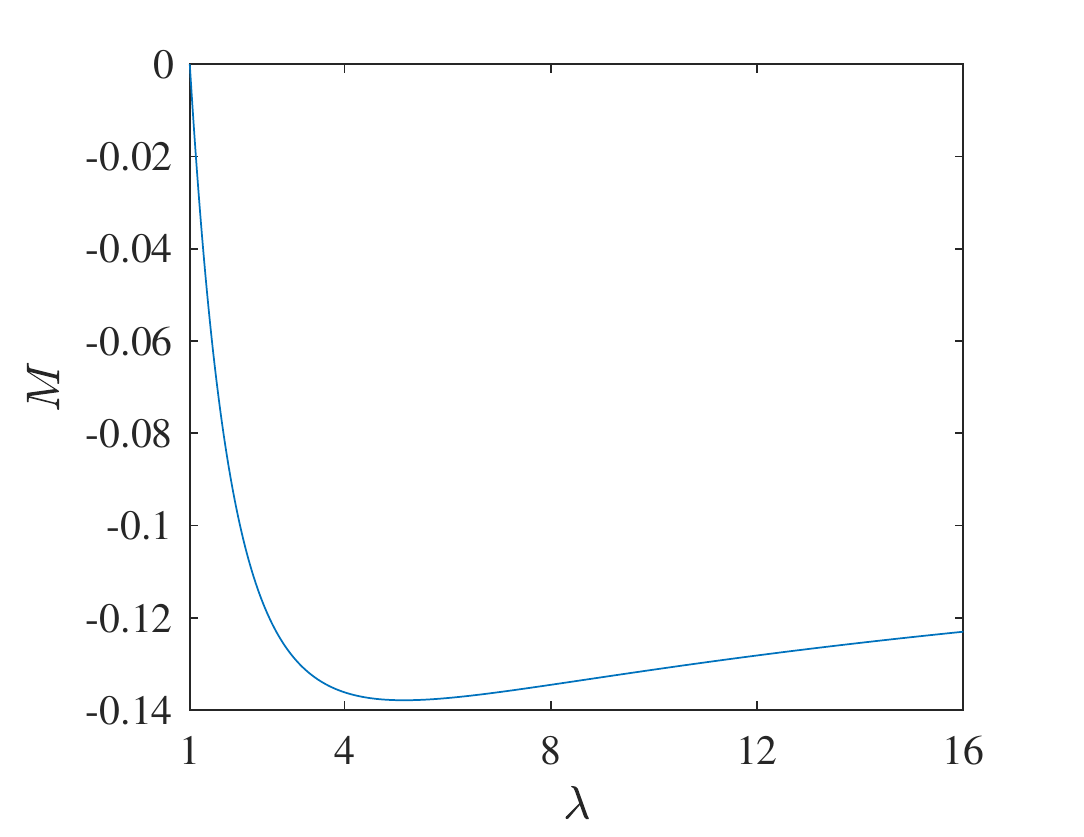}
	\caption{Shape factor $M$ as a function of aspect ratio $\lambda$.}
	\label{figShape}
\end{figure}

Using Eqs.~(\ref{eom}), we obtain the governing equation of the angular velocity of a settling swimmer (Eq.~(\ref{eqFIswimmer})). The shape parameter $M$ in Eq.~(\ref{eqFI}) is only a function of aspect ratio $\lambda$ (Figure \ref{figShape}), and is defined by $M=\mathcal{A}I_{\perp}/C_{\perp}$, where $I_{\perp}$ and $C_{\perp}$ for elongated spheroids are as follows \cite{Gustavsson2019}:
\begin{subequations}
	\begin{equation}
		I_{\perp} = \frac{1+\lambda^2}{5}a^2,
	\end{equation}
	\begin{equation}
		C_{\perp}=\frac{8 a^{2}\left(\lambda^{4}-1\right)}
		{9 \lambda\left[\left(2 \lambda^{2}-1\right) \beta-1\right]},
	\end{equation}
	\begin{equation}
		\text { with } \beta=\frac{\ln \left(\lambda+\sqrt{\lambda^{2}-1}\right)}{\lambda \sqrt{\lambda^{2}-1}},
	\end{equation}
	For elongated spheroids:
	\begin{equation}
		\mathcal{A}=\frac{5}{6 \pi} F_{\beta} \frac{\lambda^{3}}{\lambda^{2}+1},
	\end{equation}
\end{subequations}
with $F_{\beta}$ defined as \cite{Dabade2015}:
\begin{align}	
		F_{\beta}=&
		-\frac{\pi e^{2}\left(420 e+2240 e^{3}+4249 e^{5}-2152 e^{7}\right)}
		{315\left[\left(e^{2}+1\right) \tanh ^{-1} e-e\right]^{2}\left[\left(1-3 e^{2}\right) \tanh ^{-1} e-e\right]} \\
		&+\frac{\pi e^{2}\left(420+3360 e^{2}+1890 e^{4}-1470 e^{6}\right) \tanh ^{-1} e}
		{315\left[\left(e^{2}+1\right) \tanh ^{-1} e-e\right]^{2}\left[\left(1-3 e^{2}\right) \tanh ^{-1} e-e\right]} \\		
		&-\frac{\pi e^{2}\left(1260 e-1995 e^{3}+2790 e^{5}-1995 e^{7}\right)\left(\tanh ^{-1} e\right)^{2}}
		{315\left[\left(e^{2}+1\right) \tanh ^{-1} e-e\right]^{2}\left[\left(1-3 e^{2}\right) \tanh ^{-1} e-e\right]}, \label{eqFb}
\end{align}
where $e=\sqrt{1-\lambda^{-2}}$. For more details on these parameters, readers can refer to Refs. \cite{Sheikh2020,Gustavsson2019,Dabade2015}.

\section{Appendix B. Direct numerical simulation of turbulence and swimmers}
We use an Eulerian-Lagrangian method to simulate swimmers in homogeneous isotropic turbulence. The dynamics of fluid phase is resolved in an Eulerian frame, while each individual swimmer is tracked along the Lagrangian trajectory using local instantaneous flow information at swimmer position. The incompressible  turbulence is directly simulated by solving the Navier-Stokes equations:

\begin{equation}
	\frac{\partial \ve{u}}{\partial t} + \ve{u}\cdot \nabla \ve{u} = -\frac{\nabla p_{\rm f}}{\rho} 
	+ \gamma\nabla^2\ve{u} + \ve{f},
	\label{eqNS1}
\end{equation}
\begin{equation}
	\nabla \cdot \ve{u} = 0,
	\label{eqNS2}
\end{equation}
where $t$ is the time, $\ve{u}$ is the fluid velocity. The symbols $p_{\rm f}$ and $\rho$ denote the pressure and density of fluid, respectively. An external force $\ve{f}$ is applied to the large scales and injects energy in order to sustain turbulence and balances the rate of viscous dissipation at the Kolmogorov scale $\eta$ \cite{Machiels1997}. Three-dimensional periodic boundary conditions are applied on the boundaries of the cubic domain with a size $(2\pi)^3$. A pseudo-spectral method is used for solving the Navier-Stokes equations, and the 3/2 rule is utilized to reduce the aliasing error on the nonlinear term.
The turbulence Taylor-Reynolds number is $\mathrm{Re}_\lambda = u_{\rm rms}L_\lambda /\gamma = 60$, where $u_{\rm rms}$ is the root-mean-square velocity, $L_\lambda = u_{\rm rms} \sqrt{15\gamma \epsilon^{-1}}$.
We use $96^3$ grid points to resolve the turbulent flow fluctuation. The maximum wave number resolved is about 1.78 times greater than the Kolmogorov wave number to ensure the accuracy of resolution at small scales \cite{Pope2001}.
A random flow with an exponent energy spectrum is given as the initial flow field, and we use an explicit second-order Adams-Bashforth scheme for time integration of Eqs. (\ref{eqNS1}) and (\ref{eqNS2}) with a time step smaller than 0.01$\tau_{\eta}$ \cite{Rogallo1981}. 

After turbulence is fully developed, swimmers are released in the flow field with random positions and orientations. 
Fluid velocity and its gradients in Eqs. (2) and (4) are interpolated by a second-order Lagrangian method at the particle position, using fluid information at Eulerian grid points. Eqs. (1) and (2) are integrated by a second-order Adams-Bashforth scheme similar to time integration of the fluid phase. The number of particles is 120,000 for each parameter configuration, and the statistics in turbulence are obtained by averaging over more than 40 uncorrelated time samples after the statistics has reached a steady state.

\begin{table*}
	\footnotesize
	\caption{Parameters of typical plankton species \cite{Smayda2010,Sohn2011,Titelman2001,Titelman2003,Carlotti2007}.
		Data are mean values $\pm$ standard deviations. — : unavailable data. Superscript$^1$: $V_{\rm settle}$ is calculated using Stokes settling velocity assuming that the density of \emph{Cochlodinium polykrikoides} is 5.9\% greater than fluid density \cite{Kamykowski1992}.
	}
	
	\begin{tabular*}{\textwidth}{@{\extracolsep{\fill}}lllllll}

		\toprule		
		Species & &Width($\SI{}{\micro\meter}$) & Length($\SI{}{\micro\meter}$) & $\lambda$ & $v_{\rm swim}$($\SI{}{\micro\meter/\second}$) & $V_{\rm settle}$($\SI{}{\micro\meter/\second}$)\\
		
		\midrule
		Cochlodinium polykrikoides \cite{Sohn2011} 
		& single cell      & 25.1 $\pm$ 2.7     & 40.8 $\pm$ 2.0     & 1.63 $\pm$ 0.25 & 391 $\pm$ 92   & 26$^1$       \\
		& 2-cells          & 25.3 $\pm$ 1.8     & 50.7 $\pm$ 0.9     & 2.00 $\pm$ 0.18 & 599 $\pm$ 126  & 29$^1$       \\
		& 4-cells          & 25.5 $\pm$ 0.7     & 102.3 $\pm$ 4.2    & 4.01 $\pm$ 0.27 & 800 $\pm$ 129  & 42$^1$       \\
		& 8-cells          & 29.0 $\pm$ 1.4     & 182.0 $\pm$   10.9 & 6.28 $\pm$ 0.68 & 856 $\pm$ 108  & 65$^1$       \\
		Centropages typicus \cite{Titelman2003,Carlotti2007}
		& early   nauplius & 57.0 $\pm$ 11.5    & 132.0 $\pm$   16.0 & 2.31 $\pm$ 0.19 & 330 $\pm$ 210  & 50 $\pm$ 40   \\
		& late   nauplius  & 97.2 $\pm$ 22.1    & 225.0 $\pm$   33.0 & 2.31 $\pm$ 0.19 & 720 $\pm$ 310  & 140 $\pm$ 70  \\
		Euterpina acutifrons \cite{Titelman2003}
		& late   nauplius  & 86.6 $\pm$ 18.7    & 200.0 $\pm$ 27.0   & —           & 1080 $\pm$ 310 & 260 $\pm$ 50  \\
		Eurytemora affinis \cite{Titelman2003}
		& late   nauplius  & 87.4 $\pm$ 18.7    & 202.0 $\pm$   27.0 & —           & 1640 $\pm$ 400 & 182       \\
		Temora longicornis  \cite{Titelman2001,Titelman2003}
		& late   nauplius  & 133.3 $\pm$   26.3 & 308.0 $\pm$   36.0 & —           & 570 $\pm$ 140  & 240 $\pm$ 70  \\
		& copepod          & 129.0 $\pm$   26.8 & 298.0 $\pm$   38.0 & —           & 820 $\pm$ 180  & 170 $\pm$ 240 \\
		Ceratium tripos  \cite{Smayda2010}
		&                  & —              & 73.5           & —           & 167        & 164       \\
		Ceratium furca \cite{Smayda2010}
		&                  & —              & 45.1           & —           & 780        & 62        \\
		Akashiwo sanguinea \cite{Smayda2010}
		&                  & —              & 42.2           & —           & 300        & 54        \\
		Dinophysis acuminata \cite{Smayda2010}
		&                  & —              & 32.4           & —           & 332        & 32        \\
		Alexandrium minutum  \cite{Smayda2010} 
		&                  & —              & 18.1           & —           & 278        & 10        \\
		Prorocentrum minimum \cite{Smayda2010}
		&                  & —              & 12.7           & —           & 206        & 5        \\
		\bottomrule
	\end{tabular*}
	\label{table1}		
	
\end{table*}
\setlength{\intextsep }{0cm}	

\begin{table*}
	\footnotesize
	\caption{Dimensionless numbers of typical plankton species shown in Table \ref{table1}. The Kolmogorov scales of ocean turbulence is calculated with $\gamma = 1.058\times10^{-6} \SI{}{\meter^2\second^{-1}}$, and the energy dissipation rate $\epsilon$ is ranging from $1\times10^{-9}$ to $1\times10^{-6} \SI{}{\meter^2\second^{-3}}$ \cite{Kiorboe1995}.
		$\mathrm{Re_p}$ is calculated with $\mathrm{Re_p} = v_{\rm swim}L/\gamma$ because $v_{\rm swim}>V_{\rm settle}$ for many species in Table \ref{table1}. Superscript$^1$: The values are calculated with $\lambda = 2.3$ similar to \emph{Centropages typicus}. Superscript$^2$: The values are calculated with $\lambda = 2.0$.}
	\begin{tabular*}{\textwidth}{@{\extracolsep{\fill}}llllllll}	
		\toprule	
		Species & & $\Phi_{\rm swim}$ & $\Phi_{\rm settle}$ & $M$ & $\mathrm{Re_p}$ & $\mathrm{St}(\times 10^{-5})$ &
		$\Psi_I$ \\
		
		\midrule
		Cochlodinium polykrikoides \cite{Sohn2011}
		& single cell    & 2.17$\sim$0.39 & 0.14$\sim$0.03 & -0.078 & 0.015 & 0.17$\sim$5.52   & 20.7$\sim$655.2 \\
		& 2-cells        & 3.32$\sim$0.59 & 0.1$\sim$0.03 & -0.101 & 0.029 & 0.22$\sim$6.95   & 9.1$\sim$287.2 \\
		& 4-cells        & 4.44$\sim$0.79 & 0.23$\sim$0.04 & -0.136 & 0.077 & 0.45$\sim$14.11   & 3.6$\sim$112.5 \\
		& 8-cells        & 4.75$\sim$0.84 & 0.36$\sim$0.06 & -0.137 & 0.147 & 0.90$\sim$28.51  & 2.1$\sim$67.4 \\
		Centropages typicus \cite{Titelman2003,Carlotti2007}
		& early nauplius & 1.83$\sim$0.33 & 0.28$\sim$0.05 & -0.114 & 0.041 & 1.29$\sim$40.78  & 8.7$\sim$274.2 \\
		& late nauplius  & 3.99$\sim$0.71 & 0.78$\sim$0.14 & -0.114 & 0.153 & 3.75$\sim$118.48  & 1.4$\sim$44.9 \\
		Euterpina acutifrons \cite{Titelman2003}      
		& late nauplius$^{1}$ & 5.99$\sim$1.06 & 1.44$\sim$0.26 & -0.114 & 0.204 & 2.96$\sim$93.61    & 0.5$\sim$16.1 \\
		Eurytemora affinis \cite{Titelman2003}        
		& late nauplius$^{1}$ & 9.09$\sim$1.62 & 1.01$\sim$0.18 & -0.114 & 0.313 & 3.02$\sim$95.49  & 0.5$\sim$15.2 \\
		Temora longicornis \cite{Titelman2001,Titelman2003}        
		& late nauplius$^{1}$ & 3.16$\sim$0.56 & 1.33$\sim$0.24 & -0.114 & 0.166 & 7.02$\sim$222.01 & 1.0$\sim$33.1 \\
		& copepod$^{1}$       & 4.55$\sim$0.81 & 0.94$\sim$0.17 & -0.114 & 0.231 & 6.57$\sim$207.83 & 1.0$\sim$32.5 \\
		Ceratium tripos$^{2}$ \cite{Smayda2010}           
		&                & 0.93$\sim$0.16 & 0.91$\sim$0.16 & -0.101 & 0.012 & 0.46$\sim$14.60  & 5.9$\sim$186.1 \\
		Ceratium furca$^{2}$ \cite{Smayda2010}          
		&                & 4.32$\sim$0.77 & 0.34$\sim$0.06 & -0.101 & 0.033 & 0.17$\sim$5.50   & 3.3$\sim$105.9 \\
		Akashiwo sanguinea$^{2}$ \cite{Smayda2010}       
		&                & 1.66$\sim$0.30 & 0.30$\sim$0.05 & -0.101 & 0.012 & 0.15$\sim$4.81   & 9.9$\sim$314.3 \\
		Dinophysis acuminata$^{2}$ \cite{Smayda2010}      
		&                & 1.84$\sim$0.33 & 0.18$\sim$0.03 & -0.101 & 0.010 & 0.09$\sim$2.84   & 15.2$\sim$481.8 \\
		Alexandrium minutum$^{2}$ \cite{Smayda2010}       
		&                & 1.54$\sim$0.27 & 0.05$\sim$0.01 & -0.101 & 0.005 & 0.03$\sim$0.89   & 58.3$\sim$1843.6 \\
		Prorocentrum minimum$^{2}$ \cite{Smayda2010}      
		&                & 1.14$\sim$0.20 & 0.03$\sim$0.00 & -0.101 & 0.002 & 0.01$\sim$0.44   & 159.8$\sim$5053.4\\
		\bottomrule
	\end{tabular*}
	\label{table2}
	
\end{table*}

\section{Appendix C. Typical parameters of plankton}
\label{section4}	
Here, we summarize the parameters of typical plankton species~\cite{Smayda2010,Sohn2011,Titelman2001,Titelman2003,Carlotti2007}. In Table \ref{table1} we show the typical length, aspect ratio, swimming and settling speeds. In Table \ref{table2} we summarize the non-dimensional parameters. The $\Rep$ and $\rm{St}$ of these typical species are negligibly small, so the model derived in Appendix A is applicable. We also estimate the reorientation parameter $\Psi_I$ with Eq.~(\ref{eqpsiI}). $\Psi_I$ is small for fast-swimming species in turbulence with small energy dissipation rate, indicating that the effect of fluid inertial torque is significant.

\bibliography{ref}

\end{document}